\documentclass[preprint,10pt]{elsarticle}

\usepackage{amssymb}

\usepackage{pgfplots}
\usepackage{pstricks,pst-node, pst-fun}
\usepackage{dsfont}
\usepackage{amssymb}
\usepackage[british,UKenglish]{babel}
\usepackage[utf8]{inputenc}
\usepackage{amsmath}
\usepackage{relsize}
\usepackage{setspace}
\usepackage{multirow}
\usepackage{comment}
\usepackage{subcaption}
\usepackage{float}
\usepackage{algorithm}
\usepackage{algorithmic}

\usepackage{afterpage}

\journal{Journal Name}

\begin{document}

\begin{frontmatter}




\title{Multi-scale energy homogenization for 3D printed microstructures with a Diritchlet boundary condition relaxation under plastic deformation.}



\author{Antonio Tabanera$^a$, Luis Saucedo-Mora$^{a,b,c}$, Miguel Angel Sanz$^a$\\ and Francisco J. Montans$^{a,d}$}


\address{$^a$ E.T.S. de Ingeniería Aeronáutica y del Espacio, Universidad Politécnica de Madrid, Pza. Cardenal Cisneros 3, 28040, Madrid, Spain\\

\vspace{0.3cm}

$^b$ Department of Materials, University of Oxford, Parks Road, Oxford, OX1 3PJ, UK\\

\vspace{0.3cm}

$^c$ Department of Nuclear Science and Engineering, Massachusetts Institute of Technology,  MA02139, USA\\

\vspace{0.3cm}

$^d$ Department of Mechanical and Aerospace Engineering, Herbert Wertheim College of Engineering, University of Florida, FL32611, USA}


\end{frontmatter}


\section{Methodology and results} 

Metamaterials are capable to mimic a large variety of mechanical properties, such as high strength, ductility, or mitigation of low-frequency vibration \cite{mizukami2023design}, among others. All are huge improvements in the performance of the components, and are in fact, changing how the structural designs are done. Those capabilities have been widely studied experimentally \cite{neelam2022mechanical}, and its modelling is an emerging methodology that have evolved drastically in the last years. This interest have been driven by the best cost-effort relation of virtual testing with accurate results. In this work we explore a methodology to reduce the memory consumed by those calculations with a multiscale model with adjustable Dirichlet boundary conditions at the metamaterial level. It consists in two layers, one at a low scale (metamaterial) and another at a large scale (FEM of the component). We propose a methodology for the transfer and adjustment of information between layers. First, at the macromechanical level, the geometry of the sample and the boundary conditions are applied to be calculated in a FEM with solid elements. Then, the nodal displacements are used as boundary conditions of the metamaterial, where the strain energy is calculated and used for the adjustment of the mechanical properties of the FEM element. With this, the behaviour of the FEM and the metamaterial are mimetic. It also permits to make this transfer locally for every finite element and the piece of metamaterial inside it, with a local division of the model and a save of memory.

\begin{figure}[H]
\centering
\includegraphics[width=1.0\textwidth]{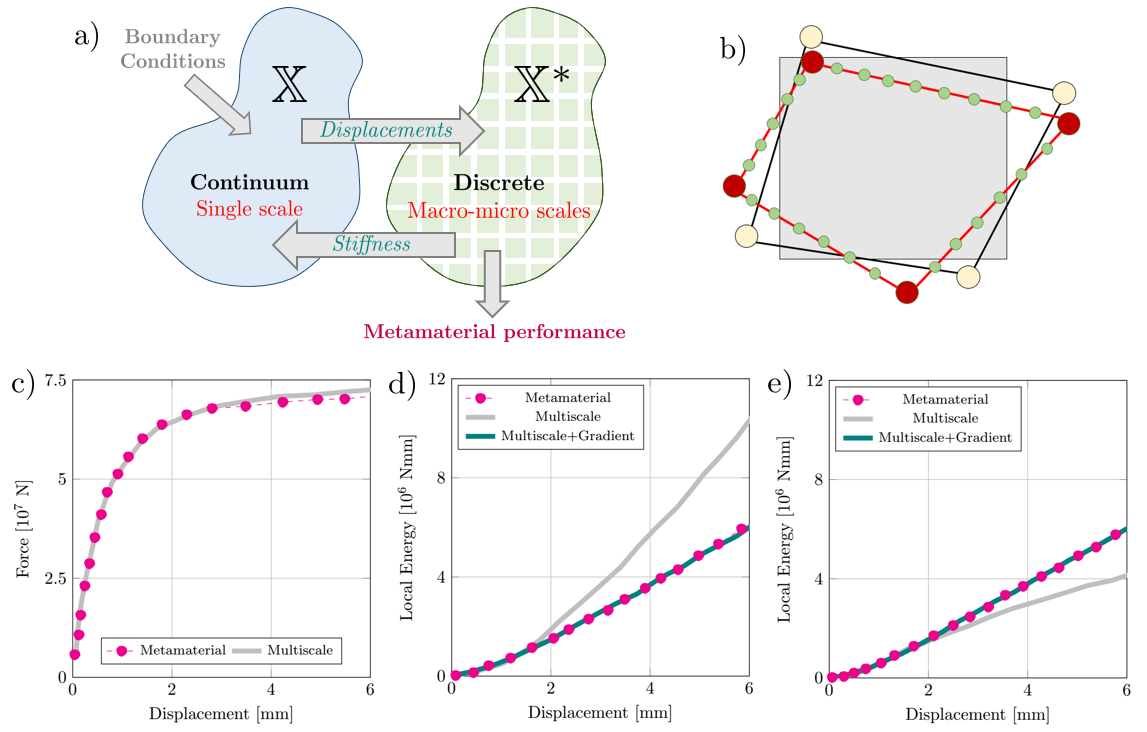}
\caption{Different phases and results of the model proposed. a) is a scheme of the information transfer between layers. On the left the FE layer where the geometry, loads and restriction of the sample are modelled, then the continuum model is calculated and the information (displacements of the nodes) is transferred into the metamaterial, right of a). Then, the stiffness of the metamaterial region within the FE analysed is transferred into the large scale mechanical model. b) shows the finite element without deformation (gray), the displacements calculated by the FEM (yellow dots), and the final displacement adjusted with the gradient method to relax the boundary conditions (red points, with the nodes of the crossing points of the metamaterial and the element face in green). c) shows the load-displacement curve of the calculation with the full scale metamaterial, and the multiscale approach without relaxation of boundary conditions through the gradient method proposed. d) and e) show, for the same calculation as c), the influence of the gradient method for the local correction of the multiscale model in comparison with the raw result and the real local behaviour of the metamaterial full scale model. Those results are for 2 different elements in the FE mesh with a more complex local conditions and a higher difference between the raw multiscale and the reference result of the metamaterial.}
\label{figresult}
\end{figure}

Full scale virtual testing in metamaterials is a complex task because of the large number of degrees of freedom that have to be solved each time. This is due to the fact that a metamaterial is made of thousands or millions of small beams ensambled together, so every beam has to be calculated. And many efforts have been made in the last years to improve the performance of those calculations \cite{jamshidian2020multiscale} \cite{phlipot2019quasicontinuum} \cite{glaesener2020continuum}  \cite{fish2021mesoscopic}. All those cited works follows the same philosophy presented here to detect which regions need to be solved with more precision to ensure good results, therefore avoiding to perform precise calculations where it is not needed. But this work presents and advantage with the gradient local calculation, which fits the real behaviour at every local region of the sample.

In order to reduce the RAM usage, the model works by splitting the large complete problem of a metamaterial structure into various smaller problems (regions) with much less degrees of freedom each, thus reducing the computational memory required. In order to impose the adequate boundary conditions for each metamaterial region (given that now some boundaries of these smaller regions are located in the interior of the structure), firstly the structure is solved as an homogeneous solid, with averaged properties (similarly to the approach followed in \cite{tessarin2022multiscale} and \cite{georgantzinos2023multi}), so an approximation of the displacement field in each point is achieved . Once the adequate interior displacements are imposed as the new BCs, the new smaller problems can be fully solved. In order to improve the accuracy (given that the homogeneous solid has not the exact same displacement field as the real metamaterial structure does), an optimization algorithm (i.e. a gradient method) that takes into account the sum of forces in the boundary nodes is used to slightly move and adjust the nodes in the boundaries of the regions until two adjacent elements have the same value and opposite sign of the reaction forces for the BC imposed to the metamaterial. This correction does not change the global values of the whole structure by much, although locally the precision increases noticeably, as Figure \ref{figresult}d and e shows.

\subsection{Computational cost reduction}

The model presented has shown major improvements in RAM consumption over the full scale metamaterial calculation where the whole structure is analysed at the same time. To prove the method, virtual tensile tests on cubic samples have been performed. With a cubic sample of only 16 unit cells per side (table \ref{case 2: 16x16x16 metamaterial cells}), the RAM usage has passed from 9.1 GB to 0.7 GB, which implies a reduction of more than 93\%, that shows the good performance of the model even with the first results with a low number of metamaterial cells. It is expected that the accuracy will increase with a higher number of metamaterial unit cells, as it is considered in \cite{mirzakhani2023effects}. Those results are for a model run in serial, but since every element of the FE mesh is calculated separately, the paralellization is direct.

\begin{table}[H]
\centering
\begin{tabular}{||c || c| c || c| c || } 
 \hline
  16x16x16 & RAM & $\Delta$ \% & $t_0$ & $\Delta$ \% \\
 \hline\hline
 metamaterial & 9.1 GB & [-] & 540 s& [-] \\
 \hline
 multiscale & 0.7 GB & -93 &450 s& -16 \\
 \hline
 multiscale gradient & 0.7 GB & -93 &900 s& +317 \\
 
 \hline
\end{tabular}
\caption{Tensile test without parallelization: 16x16x16 metamaterial cells}
\label{case 2: 16x16x16 metamaterial cells}
\end{table}

\section{Conclusions}

The present work is a proof of concept of the capabilities of paralellization in the calculation of metamaterials in a non-linear regime. In this work we subdivided the bulk material into subregions where the mechanical properties are homogenized energetically. We demonstrate that the calculation can be subdivided to save RAM memory and fit the local non-linear behaviour of the metamaterial. This methodology has the potentiality to be implemented in the parallelization of those calculations, where the right estimation of the energy of the local processes at every step is important.

\section*{Funding}
\begin{minipage}{0.15\linewidth}
\includegraphics[height=2.5\baselineskip]{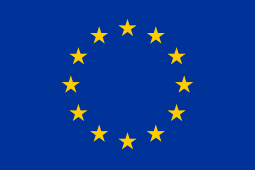}
\end{minipage}
\begin{minipage}{0.7\linewidth}
\textit{This project has received funding from the European Union's Horizon 2020 research and innovation programme under the Marie S\l{}odowska-Curie Grant Agreement No. 101007815}
\end{minipage}

\vspace{0.6cm}

\bibliographystyle{plain}
\bibliography{references}

\begin{thebibliography}{1}

\bibitem{fish2021mesoscopic}
Jacob Fish, Gregory~J Wagner, and Sinan Keten.
\newblock Mesoscopic and multiscale modelling in materials.
\newblock {\em Nature materials}, 20(6):774--786, 2021.

\bibitem{georgantzinos2023multi}
Stelios~K Georgantzinos, Panagiotis~A Antoniou, and Christos Spitas.
\newblock A multi-scale computational framework for the hygro-thermo-mechanical
  analysis of laminated composite structures with carbon nanotube inclusions.
\newblock {\em Results in Engineering}, 17:100904, 2023.

\bibitem{glaesener2020continuum}
Rapha{\"e}l~N Glaesener, Erik~A Tr{\"a}ff, Bastian Telgen, Renato~M Canonica,
  and Dennis~M Kochmann.
\newblock Continuum representation of nonlinear three-dimensional periodic
  truss networks by on-the-fly homogenization.
\newblock {\em International Journal of Solids and Structures}, 206:101--113,
  2020.

\bibitem{jamshidian2020multiscale}
Mostafa Jamshidian, Narasimha Boddeti, David~W Rosen, and Oliver Weeger.
\newblock Multiscale modelling of soft lattice metamaterials: Micromechanical
  nonlinear buckling analysis, experimental verification, and macroscale
  constitutive behaviour.
\newblock {\em International Journal of Mechanical Sciences}, 188:105956, 2020.

\bibitem{mirzakhani2023effects}
Amin Mirzakhani and Ahmad Assempour.
\newblock The effects of microstructural parameters on the tension-compression
  mechanical behavior of extruded mg-xy rods using crystal plasticity finite
  element modeling.
\newblock {\em Results in Engineering}, 17:100834, 2023.

\bibitem{mizukami2023design}
Koichi Mizukami, Rikuto Imanishi, Hitoshi Matsushita, and Yoichiro Koga.
\newblock Design and additive manufacturing of elastic metamaterials with 3d
  lever-type inertial amplification mechanisms for mitigation of low-frequency
  vibration.
\newblock {\em Results in Engineering}, 20:101389, 2023.

\bibitem{neelam2022mechanical}
Rajat Neelam, Shrirang~Ambaji Kulkarni, HS~Bharath, Satvasheel Powar, and
  Mrityunjay Doddamani.
\newblock Mechanical response of additively manufactured foam: A machine
  learning approach.
\newblock {\em Results in Engineering}, 16:100801, 2022.

\bibitem{phlipot2019quasicontinuum}
Gregory~P Phlipot and Dennis~M Kochmann.
\newblock A quasicontinuum theory for the nonlinear mechanical response of
  general periodic truss lattices.
\newblock {\em Journal of the Mechanics and Physics of Solids}, 124:758--780,
  2019.

\bibitem{tessarin2022multiscale}
Andrea Tessarin, Mirco Zaccariotto, Ugo Galvanetto, and Domenico Stocchi.
\newblock A multiscale numerical homogenization-based method for the prediction
  of elastic properties of components produced with the fused deposition
  modelling process.
\newblock {\em Results in Engineering}, 14:100409, 2022.

\end{thebibliography}

\end{document}